\documentclass[12pt]{elsart}
\usepackage{amsmath}
\usepackage{amsfonts}
\usepackage{mathrsfs}
\usepackage{graphicx}

\def\bc{\begin{center}}
\def\ec{\end{center}}
\def\om{\omega}
\def\be{\begin{eqnarray}}
\def\ee{\end{eqnarray}}
\def\prt{\partial}

\begin{document}
\begin{frontmatter}
\title{Spectra and femtoscopic scales in a hydrokinetic model for baryon-rich fireballs}
\author[BITP]{Iu.A.~Karpenko},
\author[JINR,Mold,GSI]{A.S.~Khvorostukhin},
\author[BITP,EMMI]{Yu.M.~Sinyukov},
\author[JINR,GSI]{V.D.~Toneev}
\address[BITP]{Bogolyubov Institute of Theoretical Physics, 03680 
Kiev-143, Ukraine}
\address[JINR]{Joint Institute for Nuclear Research,
 141980 Dubna, Moscow Region, Russia}
\address[Mold]{ Institute of Applied Physics, Moldova AS,
MD-2028 Kishineu, Moldova}
\address[EMMI]{ExtreMe Matter Institute EMMI, GSI Helmholtzzentrum
f\"ur Schwerionenforschung GmbH,  64291 Darmstadt,
Germany}
\address[GSI]{GSI Helmholtzzentrum
f\"ur Schwerionenforschung GmbH, D-64291 Darmstadt, Germany}

\maketitle

\begin{abstract}The hydrokinetic model and the scheme of dynamical freeze out
proposed earlier for the RHIC energy are extended to lower colliding energy
for non-vanishing baryon chemical potential. In this case a new two-phase
equation of state is applied where baryon-rich hadronic matter is
described within modified relativistic mean-field theory. As an example,
the approach is employed for analyzing  pion and kaon spectra and 
interferometry radii of Pb+Pb
collisions at laboratory energies 40 and 158 AGeV. A good agreement is
observed for the transverse mass dependence of the longitudinal radius for
both energies studied. The transverse radii $R_{out}, R_{side}$ reproduce
experiment at $E_{lab}=$158\ AGeV but at $E_{lab}=$40\ AGeV there is a
marked (20 \%) difference from experimental points. This  discrepancy is 
more noticeable in analysis of the ratio $R_{out}/R_{side}$. Nevertheless, 
considered examples of application of the hydrokinetic model with equation 
of state including the first order phase transition are quite successful 
and further development of the model is required.
\end{abstract}
\end{frontmatter}

 \section{Introduction}

 An analysis of the experimental data in nucleus-nucleus collisions obtained at Relativistic Heavy Ion Collider(RHIC) \cite{rhic-hydro}
 and very recently at Large Hadron Collider (LHC) \cite{lhc-v2} denotes hydrodynamics as an adequate tool to describe a behavior
 of the bulk of the matter created at these collisions. The reason
 could be that the collective behavior of the quark-gluon plasma at the reached temperatures is similar to what an almost
 perfect fluid has. The hydrodynamic models use then the thermodynamic equation of state (EoS) reconciled with the Lattice QCD. The successful
description of RHIC data within hydrodynamic approaches with
artificial sharp freeze-out \cite{cracow} as well as with dynamical
 freeze-out in the HydroKinetic Model (HKM)~\cite{PRL,PRC,PRC1} 
were done and
the predictions for the spectra and femtoscales at LHC are in
quite reasonable agreement \cite{lhc-v2,cracow,PLB}.

    As for the applicability of hydrodynamic approach to lower
 energies, such as the SPS ones and planning experiments
 NICA and FAIR, it is a questionable issue. First, as expected,
 created fireballs spend only small part of its lifetime if any in
 quark-gluon phase. An evolution of the hadronic matter could be
 non-equilibrated, neither thermally, nor chemically. Since in the HKM
model~\cite{PRL} the evolution and gradual decay of
 the hadronic matter are described as non-equilibrated process, it seems
 reasonable to try to apply this model to relatively small collision
energies. It is the main aim of our work.

  One of the crucial points is the equation of state
 at large baryonic chemical
 potentials, which are typical for these energies. As known, because of the
 principal technical problem the EoS cannot
 be found from the lattice QCD calculations at large baryonic chemical
 potentials. Currently such an estimate should be obtained rather from a 
 detailed study of strongly interacting hadronic matter supplemented by the
model of quark-gluon phase at large $\mu_B$ and $\mu_S$. In this
paper we will use the EoS taking into account a possible phase transition
between quarks-gluons and hadrons including in-medium modification of  hadron properties in hot and dense nuclear surrounding~\cite{KTV07}

\section{Basic features of the model }
 Let us briefly describe the main features of the HKM \cite{PRC,PRC1}.
It incorporates hydrodynamical expansion of the systems formed in
\textit{A}+\textit{A} collisions and their dynamical decoupling
described by escape probabilities. The basic hydro-kinetic code
includes decays of resonances into expanding hadronic
chemically non-equilibrated system and, based on the resulting
composition of the hadron-resonance gas at each space-time point,
provides the equation of state (EoS) in a vicinity of this
point. The obtained local EoS allows one to determine the further
evolution of the considered fluid elements. The complete picture
of the physical processes in central Pb + Pb collisions encoded in
calculations is the following.

\subsection{Initial conditions}
 Our results are all related to the
central rapidity slice where we use the boost-invariant
Bjorken-like initial condition. We consider the proper time of
thermalization of quark-gluon matter as  the minimal one discussed
in the literature, $\tau_0=1$ fm/c. The initial energy density in
the transverse plane is supposed to be Glauber-like \cite{Kolb},
i.e. is proportional to the participant nucleon density  for Pb+Pb
(SPS) and Au+Au (RHIC, LHC) collisions with zero impact parameter.
The height of the distribution - the maximal initial energy
density  $\epsilon(r=0)=\epsilon_0$ is a fitting parameter.
 Also, at time $\tau_i$, there is peripheral
region with relatively small initial energy densities:
$\epsilon(r)< 0.5$ GeV/fm$^3$. This part of the matter ("corona")
have no chance to be involved in thermalization process
\cite{Werner}. Thus, it should be considered
separately from the thermal bulk of the matter and should not
include it in hydrodynamic evolution.
 By itself the corona gives no essential
contribution to the hadron spectra for central collisions \cite{Werner}.

At the time of thermalization, $\tau_0=1$ fm/c, the system
has already developed collective transverse velocities \cite{flow,JPG}.
The initial transverse rapidity profile is supposed to be linear
in transverse radius $r_T$:
\begin{equation}
\eta_T=\alpha\frac{r_T}{R_T} \quad \text{where} \quad R_T=\sqrt{<r_T^2>}.
\label{yT}
\end{equation}
Here $\alpha$ is a second fitting parameter. Note that the
fitting parameter $\alpha$ should also include a positive
correction for underestimated resulting transverse flow since in
this work we do not take directly into account  viscosity
effects \cite{Teaney} neither at the QGP stage nor at hadronic one. In
HKM formalism~\cite{PRC} viscosity effects at the hadronic
stage are incorporated in mechanisms of the back reaction of
particle emission in hydrodynamic evolution which we ignore in
current calculations. Since the corrections to transverse flows
which depend on unknown viscosity effects are unknown, we use
a fitting parameter $\alpha$ to describe the ``additional unknown
portion'' of flows, caused by three factors: development of the
pre-thermal flows, viscosity effects in quark-gluon plasma and
in addition event-by-event fluctuations of the initial
conditions. The latter also leads to an increase of the ``effective''
transverse flows, obtained by averaging at the final stage, as
compared to the results based on the initial conditions averaged
over initial fluctuations \cite{Hama1}. Since we use the last
type of initial conditions, it should also lead to an increase of the
effective parameter $\alpha$.

\subsection{Equation of state}
 We match the equation of state at the point of the chemical freeze out $T_{ch}$ which separates chemically equilibrated and non-equilibrated thermodynamic zones.    As to the chemically equilibrated region $T>T_{ch}$ the EoS is constructed on the basis of the modified relativistic mean-field model
with Scaled Hadron Masses and Couplings (SHMC)~\cite{KTV07,KTV08}.
In this SHMC model the Lagrangian density of hadronic matter is a sum
of several terms:
 \be\label{math}  \mathcal{L}=\mathcal{L}_{\rm bar}+\mathcal{L}_
 {\rm MF}+\mathcal{L}_{\rm ex}~.
 \ee
The  Lagrangian density of the baryon component interacting via
$\sigma,\omega$ mean fields  is as follows:
 \be
 \mathcal{L}_{\rm bar} &=& \sum_{b\in {\rm \{ bar \} }} \left[ i\bar \Psi_b\,
\Big(\prt_\mu +i\,g_{\om b} \,{\chi}_\om
 \ \om_\mu \Big) \gamma^\mu\, \Psi_b -m_b^*\,
\bar\Psi_b\,\Psi_b
 \right] .
 \label{lagNn}
 \ee
  The considered baryon set $\{b\}$ consists of  all baryons and 
low-lying resonances with the mass
$m_b<$1700 MeV, including strange and double-strange hyperons as well as
antiparticles. The used $\sigma$-field dependent effective masses
of baryons are~\cite{KTV07,KTV08,KV04}
\be \label{bar-m}
{m_b^*}/{m_b}=\Phi_b(\chi_\sigma  \sigma)= 1 -g_{\sigma b} \
\chi_{\sigma} \ \sigma /m_b \,, \; b\in\{b\}~.
 \ee
 In Eqs.
(\ref{lagNn}), (\ref{bar-m})  $g_{\sigma b}$ and $g_{\om b}$  are
coupling constants and $\chi_\sigma (\sigma)$, $\chi_\om (\sigma)$
are coupling  scaling functions.

The $\sigma$-, $\omega$-meson mean field contribution is given by
\begin{eqnarray} \mathcal{L}_{\rm
MF}&=&\frac{\prt^\mu \sigma \ \prt_\mu
\sigma}{2}-\frac{m_\sigma^{*2}\,
\sigma^2}{2}-{U}(\chi_{\sigma}\sigma)
-\frac{\omega_{\mu\nu}\,\omega^{\mu\nu}}{4} +\frac{m_\om^{*2}\,
\om_\mu\om^\mu}{2}~,\\
 \omega_{\mu\nu}\,&=&\partial_\mu \om_\nu -\partial_\nu \om_\mu ~,\quad U(\chi_{\sigma}\sigma)
 =m_N^4 (\frac{b}{3}\,f^3 +\frac{c}{4}\,f^4 ),
 \quad  f=g_{\sigma N} \ \chi_\sigma \ \sigma/m_N\,.\nonumber
\end{eqnarray}
 There exist only $\sigma$ and $\om_0$ mean field solutions of
equations of motion. The mass terms of the mean fields are \be
\label{bar-m1} {m_m^*}/{m_m}&=&|\Phi_m (\chi_\sigma \sigma)|\,,
\quad \{m\}=\sigma,\om\,. \ee

The dimensionless scaling functions $\Phi_b$ and $\Phi_m$, as well
as the coupling scaling functions $\chi_m$, depend on the scalar
field in combination $\chi_\sigma(\sigma) \ \sigma$.
 Following
\cite{KV04} we assume approximate validity of the Brown-Rho
scaling ansatz in the simplest form
\be \label{Br-sc}\Phi =\Phi_N
=\Phi_\sigma =\Phi_\om =\Phi_\rho =
 1-f .
 \ee
The third term in  the Lagrangian density (\ref{math}) corresponds to
meson quasiparticle excitations, where particles with $m_m<$ 1100 MeV
are included
 The choice of parameters and other details of the SHMC model
can be found in~\cite{KTV07,KTV08}.

\begin{figure}[htb]
\begin{center}
 \includegraphics[scale=0.45]{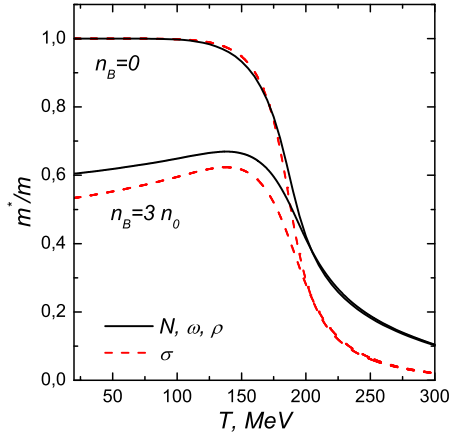}
 \caption{The temperature dependence of effective  masses of
the nucleon, $\omega$ and $\rho$ excitations (solid line) and of
the $\sigma$-meson excitation (dashed line) calculated within the
SHMC model for two values of  the baryon density $n_{\text B}$.}
\label{mass} \end{center}
\end{figure}
Within SHMC model  different thermodynamical
quantities in thermal equilibrium hadron matter are calculated at fixed
temperature $T$ and baryon chemical potential $\mu_{\rm B}$. In
Fig.\ref{mass} the $T$ and $\mu_{\rm B}$ dependence of hadron masses is
pictured. Note that these masses sharply decrease in the vicinity of the
expected phase transition and thereby affect on the fireball evolution.
\begin{figure}[thb]
\begin{center}
\includegraphics[width=130mm,clip]{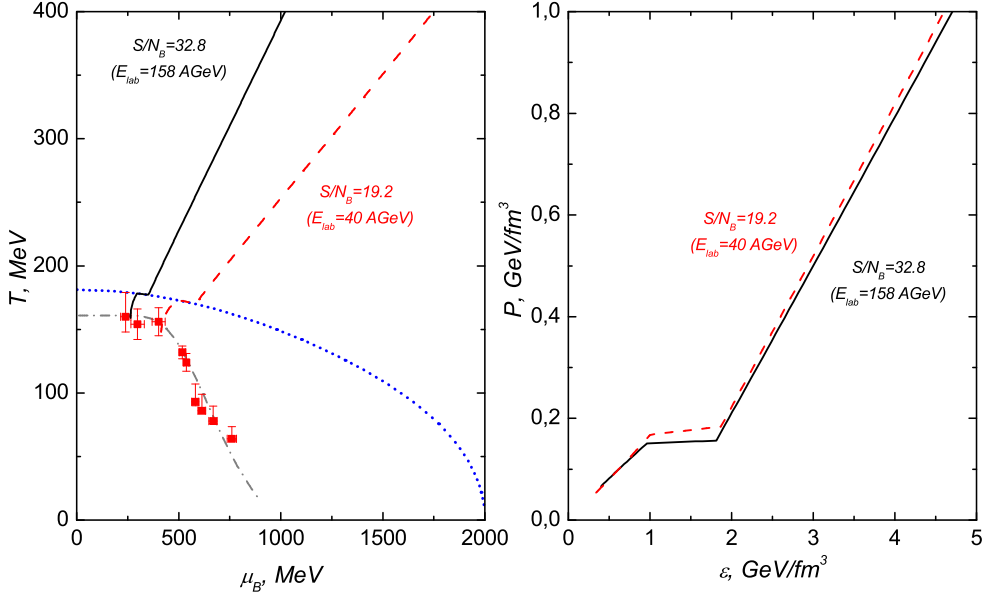}
\caption{{\it Left panel}: Isoentropic trajectories in the 
$T-\mu_{\text B}$ plane
for two values of entropy-to-baryon number ratio. The quark-hadron phase
boundary~\cite{KTV08} and the freeze out curve~\cite{PBM1} are shown
by the dotted and dashed dash-dashed lines, respectively. Experimental 
points are from review article~\cite{PBM1}.
{\it Right panel}: Equation of state within two-phase model for the same
values of the entropy baryon number ratio as in the left panel.
}
 \label{epT4} \end{center}
\end{figure}

The quark-gluon plasma phase is treated in the conventional quark bag model
with massive quark and gluons.
A two-phase model allowing for the first order phase transition from
the hadron phase (H) to the strongly coupled quark gluon plasma (Q) is
constructed by means of the Gibbs conditions. These are conditions
 for thermal ($T^{\rm Q}=T^{\rm H}$), mechanical
($P^{\rm Q}=P^{\rm H}$) and chemical ($\mu_{\rm B}^{\rm
Q} =\mu_{\rm B}^{\rm H}, \ \mu_{\rm str}^{\rm Q}
=\mu_{\rm str}^{\rm H}$) equilibrium.

There is a reasonable agreement of our two-phase thermodynamics with
available QCD lattice calculations~\cite{KTV07,KTV08}. For $\mu_B=$0
some discrepancy is seen in energy
density near the critical temperature $T_c$ due to the fact that by
construction we have the first order phase transition
while in lattice calculations we deal with a sharp crossover-type
transition. The colliding energies of interest are below an anticipated
critical point and the first order phase transition is expected.
Unfortunately, there is no lattice data for these large $\mu_B$ range. In
the left panel of Fig.\ref{epT4} we demonstrate isoentropic trajectories
for two fixed ratios of entropy $S$ to baryon number $N_B$, $S/N_B=$19.2
and 32.8,  corresponding to that of an expanding fireball at $E_{lab}=$
40 and 158 AGeV.
When a trajectory reaches the phase boundary it is flattening  in accordance
with the mixed phase of the Maxwell construction in the two-phase
model (see the right panel in Fig.\ref{epT4}). Assuming
chemical equilibrium our hydrodynamics is used till the chemical freeze
out curve shown in Fig.\ref{epT4} by the dotted line. At temperature below
$T_{ch}$ we apply non-equilibrium treatment.

%
 The EoS for baryon-rich fireballs described above is used up to
 the temperature of the chemical equilibrium $T_{ch}$.
 We should match this chemically
equilibrated region  $T>T_{ch}$ and non-equilibrated hadronic
region at $T<T_{ch}$. The last is described in the HKM as an ideal gas
of hadrons and their resonances  in
correspondence with the  concept of chemical freeze-out.
We fix the chemical freeze-out temperature and baryon chemical 
potential to be
$\{T,\mu_\text{B}\}=\{160 \text{ MeV}, 220 \text{ MeV}\}$ for
$E_\text{Lab}=158$ GeV collisions and
$\{T,\mu_\text{B}\}=\{148 \text{ MeV}, 370 \text{ MeV}\}$ for
$E_\text{Lab}=40$ GeV collisions,
based on $T_{ch}(\sqrt{s_{NN}})$, $\mu_\text{B}(\sqrt{s_{NN}})$
parameterizations obtained
from particle number ratio analysis performed in \cite{andronic}. In both
cases, strange chemical potential is calculated from the condition of
zero strange density.
 The proper choice of $T_{ch}$ and $\mu_{\text B}$ guarantees us the
correct particle number ratios for all quasi-stable particles
(here we calculate only pion and kaon observables). Below
$T_{ch}$ a composition of the hadron gas is changed only due to
resonance decays into expanding fluid. We include $N_h=326$
well-established hadron states made of $u$, $d$, $s$-quarks with
masses up to 2.6 GeV from PDG table \cite{pdg},
as well as the most possible decay channels of them. The EoS
in this chemically  non-equilibrated system depends now on particle
number densities $n_i$ of all the $N_h$ particle species $i$:
$p=p(\epsilon,\{n_i\})$.  Since the energy densities in expanding
system do not directly correlate with resonance decays, all the
variables in the EoS depend on space-time points and so an
evaluation of the EoS is incorporated in the hydrodynamic code.
We calculate the EoS below $T_{ch}$ in the Boltzmann approximation of
ideal multi-component hadron gas. In addition to standard HKM
method we add now excluded volume $v_0$ (equal for all hadrons)
which may be important for baryon-rich systems, according to the
following formula:
\begin{equation}\label{p-excl}
 p(T,\{\mu_i\})=\sum_i p_i^\text{id}(T,\{\mu_i-v_0\cdot p\})
\end{equation}
the superscript $id$ refers to ideal gas case. We solve this equation
numerically, and obtain other thermodynamic quantities using general
thermodynamic relations and the solution of Eq.(\ref{p-excl}).

We match energy and baryon number densities in both EoS at chemical
freeze-out point,
and minimize pressure difference there as well. We choose the excluded
volume value $v_0=0.2\ fm^3$ from the condition of the best matching.
The corresponding energy and net baryon densities are
$\{\epsilon,n_{\text B}\}=\{0.4\text{ GeV/fm$^3$},0.085
\text{ 1/fm$^3$}\}$ for
$E_\text{Lab}=158$ GeV and $\{\epsilon,n_{\text B}\}=\{0.34\text{
GeV/fm$^3$},\\ 0.12\text{ 1/fm$^3$}\}$ for $E_\text{Lab}=40$ GeV collisions,
which correspond to the chemical freeze-out parameters shown above.
The chosen excluded volume in chemically non-equilibrated ideal hadron gas
thus mimics more complicated interactions happening
in hadron matter in two-phase model before chemical freeze-out.

\subsection{Evolution}
 At the temperatures higher than $T_{ch}$ the
hydrodynamic evolution is related to the possible quark-gluon,
mixed and hadron phases which are in chemical equilibrium with
some baryonic chemical potential. The evolution is described by
the conservation law for the energy-momentum tensor of perfect
fluid:
\begin{equation}
\partial_\nu T^{\mu\nu}(x)=0~.
\label{conservation}
\end{equation}
Since in perfect hydrodynamics the entropy is conserved, the system
evolves along trajectories shown in Fig.\ref{epT4}.

At $T<T_{ch}$ the system evolves as chemically
 non-equilibrated hadronic gas. The concept of the chemical freeze-out
implies that afterwards  only elastic collisions and resonance
decays take place because of relatively small densities allied
with a fast rate of expansion at the last stage. Thus, in addition
to Eq.(\ref{conservation}), the equations accounting for the particle
number conservation and resonance decays are added. If one
neglects the thermal motion of heavy resonances, the equations for
particle densities $n_i(x)$ take the form:
\begin{equation}
\partial_\mu(n_i(x) u^\mu(x))=-\Gamma_i n_i(x) + \sum\limits_j
 b_{ij}\Gamma_j  n_j(x)
\label{paricle_number_conservation}
\end{equation}
where $b_{ij}=B_{ij}N_{ij}$ denote the average number of i-th
particles coming from arbitrary decay of j-th resonance,
$B_{ij}=\Gamma_{ij}/\Gamma_{j,tot}$ is branching ratio, $N_{ij}$
is a number of i-th particles produced in $j\rightarrow i$ decay
channel.  Eqs.(\ref{conservation}) and
Eqs.(\ref{paricle_number_conservation}) are solving simultaneously
with calculation of the  EoS, $p(x)=p(\epsilon(x),\{n_i(x)\})$ at
each point $x$.

\subsection{System's decoupling and spectra formation} During the
matter evolution, in fact, at $T\leq T_{ch}$, hadrons continuously
leave the system. Such a process is described by means of the
emission function $S(x,p)$ which is expressed through
the {\it gain} term for $i$-th sort of particles, $G_i(x,p)$, in
Boltzmann equations and
the escape probabilities~\cite{PRL,PRC}
\be
S_i(x,p)&=&G_i(x,p){\cal P}_i(x,p) \nonumber \\
{\cal P}_i(x,p)&=&\exp(-\int\limits_{t}^{\infty}
dsR_{i+h}(s,{\bf r}+\frac{{\bf p}}{p^0}(s-t),p))~.
\nonumber \ee
 For particle emission in the relaxation time approximation $G_{\pi}\approx
f_{\pi}R_{\pi+h}+G_{H\rightarrow\pi}$ where $f_{\pi}(x,p)$ is the
Bose-Einstein phase-space distribution, $R_{i+h}(x,p)$ is
the total collision rate of the particle, carrying momentum $p$, with
all the hadrons $h$ in the system in a vicinity of point $x$ and  the
term $G_{H\rightarrow i}$ describes an inflow of particles into
phase-space point $(x,p)$ due to the resonance decays. It is
calculated according to decay kinematics  with
simplification that the spectral function of the resonance $H$ is
$\delta(p^2-\langle m_H\rangle^2)$. The cross-sections of interacting
particles in the hadronic gas, which determine
the escape probabilities ${\cal P}(x,p)$ and emission function
$S(x,p)$ via the collision rate $R_{i+h}$, are calculated in accordance
with the UrQMD method~\cite{UrQMD}. The spectra and correlation functions
are found from the emission function $S$ in the standard way
(see, e.g.,\cite{PRL}).

\section{Results and conclusions}

\begin{figure}[b]
\hspace*{-5mm} \includegraphics[scale=0.75]{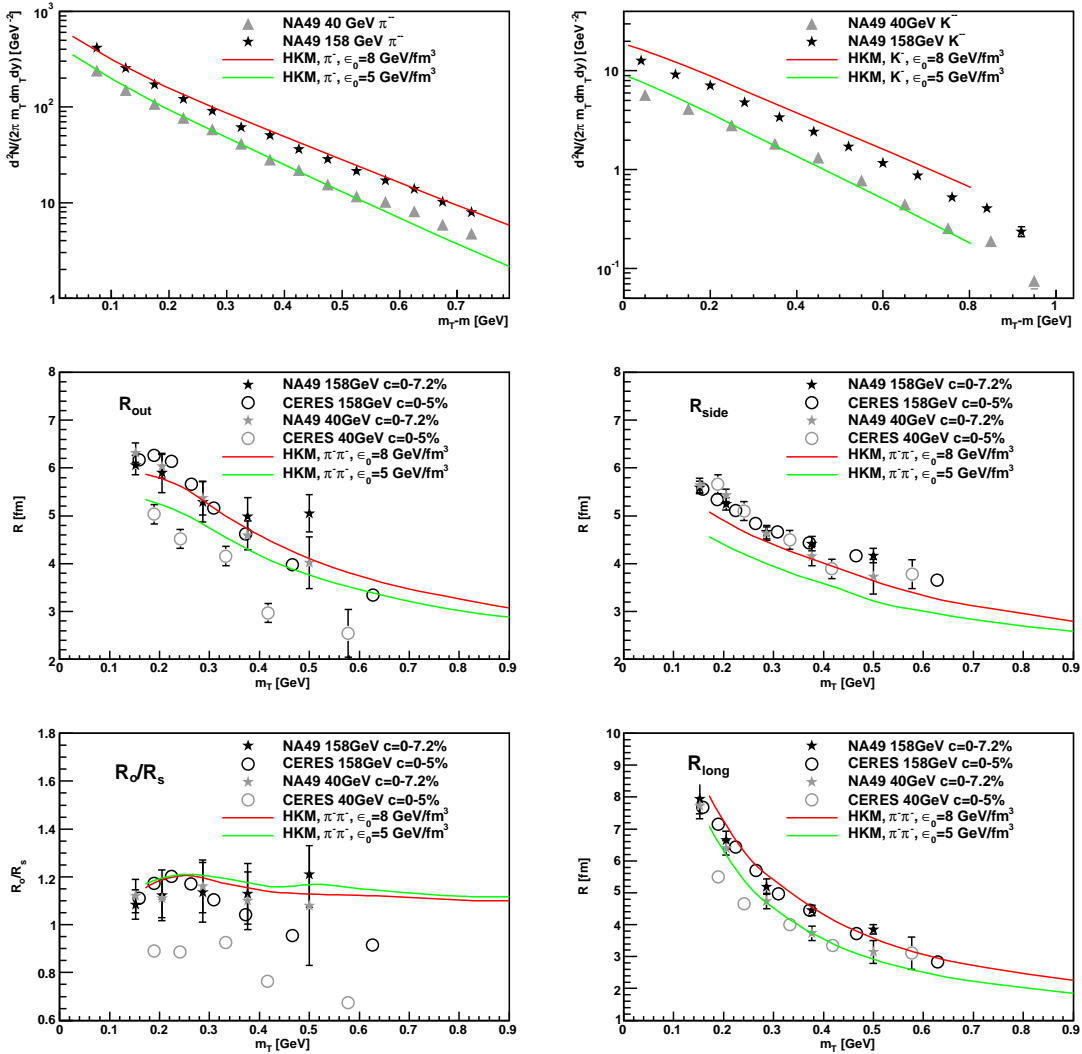}
 \caption{Pion and kaon spectra (top), pion $R_\text{out}$,
$R_\text{side}$ (middle), $R_\text{long}$ and $R_\text{out}/R_\text{side}$
ratio  (bottom) as a function of transverse momentum at the given
centrality. Experimental points for transverse spectra and interferometry
radii are taken from \cite{na49-spectra} and \cite{na49-hbt, ceres-hbt},
respectively.}\label{fig-spectra}
\end{figure}

The results for transverse mass spectra  and interferometry radii 
are presented in
Fig. \ref{fig-spectra} for the two SPS energies. The two fitting
parameters: maximal initial energy density  $\epsilon_0$ and the initial transverse velocity (associated with parameter $\alpha$) averaged over the energy profile $\langle v_T \rangle$ are equal to
$\epsilon_0=5$~ GeV/fm$^3$
($\langle\epsilon\rangle=3.5$~GeV/fm$^3$) for $E_{Lab}$=40~AGeV
and $\epsilon_0=8$~GeV/fm$^3$
($\langle\epsilon\rangle=5.6$~GeV/fm$^3$) for $E_\text{Lab}$=158
AGeV and $\langle v_T \rangle=0.196$ for both energies.  As one
can see the pion and kaon transverse mass spectra, their slopes as
well as the absolute values are in a quite reasonable agreement
with available experimental data~\cite{na49-spectra}.

 A good agreement with experiments~\cite{na49-hbt,ceres-hbt}
is also observed for the longitudinal radius $R_{long}$ for the both
energies studied. The $R_{out}$ and  $R_{side}$ interferometry radii are
well reproduced for the top SPS energy including even the ratio
$R_{out}/R_{side}$ which is more sensitive to model parameters.
For $E_{lab}=$40\ AGeV there are some deviations of the calculated curves
from experimental points which are more pronounced for the
$R_{out}/R_{side}$ ratio.

Summarizing, one can note that the hydrokinetic model can serve as
the basic dynamical model also for low-energy collisions where the
equation of state corresponding to the first order phase
transition should be used rather than that with crossover
transformation between quark-gluon and hadron matter. Such an
EoS at large chemical potentials can be successfully constructed
by matching the relativistic mean-field approximation for EoS at $T>T_{ch}$
with ideal hadronic gas with hadron masses below 2.6 GeV including
the excluded volume assumption at $T<T_{ch}$. This approach reasonably 
reproduces not only bulk properties of produced particles but also such 
delicate characteristics as interferometry radii. It is worthy to note that 
even at E$_{\text lab}$= 40\ AGeV the description of experimental data 
requires to take into account quark-gluon degrees of freedom in initial
thermodynamic conditions. The HKM  may be useful in predictions of 
collective observable in the energy considered
in future FAIR and NICA experiments.

\section*{Acknowledgments}
We are thankful to P. Braun-Munzinger and D. Voskresensky for useful 
comments. The work was supported 
in part by the State Fund for
Fundamental Researches of Ukraine, Agreement No F28/335-2009 and
Russian Fund of Fundamental Researches, Agreement Ukr-f-a
09-02-90423 for the Bilateral project SFFR (Ukraine) – RFFR
(Russia) and by the DFG grant WA 431/8-1. The research is carried 
out within the scope of the
EUREA: European Ultra Relativistic Energies Agreement (European
Research Group: Heavy ions at ultrarelativistic energies)
supported by SFFR and NASU.

\end{document}